\documentclass[12pt]{article}
\usepackage{amsfonts,amsbsy}
\usepackage{amssymb}
\usepackage{graphicx}
\usepackage{amssymb,amsbsy,url}
\newcommand{\rmi}{\mathrm{i}}
\newcommand{\fl}{\hspace*{-5mm}}

\newcommand{\be}{\begin{equation}}
\newcommand{\ee}{\end{equation}}
\newcommand{\ba}{\begin{array}}
\newcommand{\ea}{\end{array}}
\newcommand{\p}{\partial}

\newcommand{\ds}{\displaystyle}
\newtheorem{theorem}{Theorem}
\newtheorem{prop}{Proposition}
\begin{document}
\title{
\Large\bf Exact solvability of superintegrable Benenti systems
}
\author{Artur Sergyeyev\footnote{Electronic mail: Artur.Sergyeyev@math.slu.cz}\\
Silesian University in Opava, Mathematical Institute,\\
Na Rybn\'\i{}\v{c}ku 1,
746\,01 Opava, Czech Republic}

\maketitle
\begin{abstract}
We establish quantum and classical exact solvability
for two large classes of maximally superintegrable Benenti systems
in $n$ dimensions with arbitrarily large $n$.
Namely, we solve the Hamilton--Jacobi
and Schr\"odinger equations for the systems in question.
The results obtained are
illustrated for a model with the cubic potential.
\end{abstract}

\newpage




\section{Introduction}

Completely integrable systems in classical mechanics
are well known to be of great interest for both theory and
applications, see e.g. \cite{arn,bl} and references therein.
Indeed, the possibility of analytical description of the
corresponding dynamics
enables us to uncover
important physical properties of the systems in question.
The prime examples of this are the Kepler laws
and numerous physical models based on (superposition of)
harmonic oscillators.

Interestingly, 
a number of physically relevant
exactly solvable models (e.g., the Coulomb problem for the hydrogen atom and the multidimensional
harmonic oscillator, to name just a few) are {\em maximally superintegrable} rather
than just completely integrable.
In general, a Hamiltonian dynamical system
on a $2n$-dimensional phase space 
is {\em maximally superintegrable}, if it possesses the maximal
possible number, $2n-1$, of functionally independent, globally
defined integrals of motion (contrast this with $n$ {\em commuting}
integrals of motion for completely integrable systems); see e.g.\
\cite{rw83,wint04} and references therein for further details.
In this case we shall also say that the Hamiltonian of the
system in question is {\em maximally superintegrable}.
\looseness=-1

Quantizing a generic completely integrable classical system and
solving the resulting Schr\"o\-din\-ger equation may often represent
a nontrivial problem. 
However, solving the Schr\"o\-din\-ger equation
for maximally superintegrable systems is often easier
(sometimes to the extent of reducing
the determination of energy spectrum to a purely algebraic problem, as is e.g.\
the case for the nonrelativistic hydrogen atom and the
multidimensional harmonic oscillator)
because of the presence of additional integrals of motion \cite{tsw}.
\looseness=-1

These facts have lead to a considerable interest
in superintegrable systems in general, and maximally
superintegrable systems with natural Hamiltonians
on two- or three-dimensional configuration space are
now quite well understood, see e.g.\ \cite{fri65}--\cite{sw06} and
the survey \cite{wint04}. However, much less is known about
superintegrable systems in higher dimensions, although such systems
often can be interpreted as multiparticle systems consisting
of several one-, two- or three-dimensional particles with nontrivial
interactions and therefore also can be of interest in physics.
\looseness=-1

Moreover, quantum systems
on an $2n$-dimensional phase space that are exactly solvable for
arbitrary $n$
are relatively scarce, with a few beautiful exceptions
like 
the multidimensional Coulomb problem and its extensions \cite{rod02}
and the Calogero--Moser--Sutherland and related models, see e.g.\
\cite{per} and references therein. Therefore, any new examples would
be of considerable interest. Higher-dimensional superintegrable
systems are prime candidates to yield such examples: for instance,
the above examples are superintegrable, see e.g.\ \cite{rw83, rod02,
gon98} and references therein. In the present work we show that for
two large classes of maximally superintegrable Hamiltonians on an
$2n$-dimensional phase space with arbitrary $n$ 
the exact solutions for the Hamilton--Jacobi and the Schr\"odinger
equation are readily available, see Theorems~\ref{qposth} and
\ref{negkth}
below for details. \looseness=-2

More specifically, we show that solving the Hamilton--Jacobi and the
Schr\"odinger equations for the systems in question
on the $n$-dimensional configuration space with arbitrary
$n$ amounts (provided $n$ is sufficiently large)
to solving $k$-dimensional {\em
reduced\/} equations, where $k$ is fixed and independent of $n$,
as presented in Theorems~\ref{qposth} and \ref{negkth}. Moreover, the said
reduced equations can be 
readily solved,
as illustrated by the example of an anharmonic oscillator in Section
V.

In order to establish these results, we reveal a hidden symmetry of
the systems under study. This symmetry manifests itself upon passing
to the flat coordinates of the metric tensors associated with the
kinetic-energy parts of the Hamiltonians under study, as described
in Theorem~\ref{cyclcor} below. Namely, 
some of
these flat coordinates are cyclic coordinates for the Hamiltonians
under study, so the corresponding momenta are integrals of motion,
and the separation of variables is a reduced one in the sense of
\cite{benq}. The dynamics in the non-cyclic coordinates is described
by the reduced equations mentioned above. \looseness=-1

Note that the metric tensors in question have the so-called maximally
balanced signature (the numbers of plus
and minus signs differ at most by one), and the above flat coordinates
are the light-cone ones rather than the orthogonal ones.
This leads to an interesting phenomenon: the aforementioned reduced
equations are  {\em first-order} rather than second-order PDEs,
which makes them easier to solve.

\section{Preliminaries}
For fixed integer $m$ and $k$
consider the separable Benenti Hamiltonians \cite{ben93, ben97} on the phase space
$T^{\ast}\mathcal{Q}$, the cotangent space of an $n$-dimensional Riemannian manifold
$\mathcal{Q}$ endowed with the contravariant metric tensor $G_{m}$:
\begin{equation}
H_{r}^{(m,k)}=\frac{1}{2}\mu^{T}K_{r}G_{m}\mu+V_{r}^{(k)}\quad
r=1,\ldots,n.
\label{Ham}%
\end{equation}
Here $\lambda=(\lambda^{1},\ldots,\lambda^{n})^{T}$ are coordinates
on $\mathcal{Q}$ and $\mu =(\mu_{1},\ldots,\mu_{n})^{T}$ are the
corresponding momenta.
%
%
The contravariant metric
tensors $G_{m}$ have the form \cite{mac2005}\looseness=-1
\[
G_{m}=L^{m}G_{0},\quad m\in\mathbb{Z},\quad G_{0}=\mathrm{diag}\left(  \frac{1}{\Delta_{1}%
},\ldots,\frac{1}{\Delta_{n}}\right),%
\]
where $\Delta_{i}=%
{\textstyle\prod\limits_{j\neq i}}
(\lambda^{i}-\lambda^{j})$, and
$L=\mathrm{diag}(\lambda^{1},\ldots,\lambda ^{n})$ is a
$(1,1)$-tensor on $\mathcal{Q}$ called a {\em special conformal
Killing tensor} \cite{cra}.
The Killing tensors $K_{r}$ from (\ref{Ham}) are diagonal in the
$\lambda$-coordinates and can be constructed as
follows \cite{ben93}:%
\begin{equation}
K_{1}=\mathbb{I},\quad
K_{r}=\sum_{k=0}^{r-1}(-1)^{k}\sigma_{k}L^{r-1-k},\quad
r=2,\dots,n,\label{Krec}%
\end{equation}
where $\mathbb{I}$ is an $n\times n$ unit matrix, and
$\sigma_{k}=\sigma_{k}(\lambda)$ are symmetric polynomials in the
variables $\lambda^{1},\dots
,\lambda^{n}$ ($\sigma_{0}=1$, $\sigma_{1}=\sum_{i=1}^{n}\lambda^{i}%
,\dots,\sigma_{n}=\lambda^{1}\lambda^{2}\cdots\lambda^{n}$). They
are related to the coefficients of the characteristic polynomial of the
tensor $L$ as follows:
\begin{equation}
\det(\xi\mathbb{I}-L)=\sum\limits_{i=0}^{n}(-1)^{i}\sigma_{i}\xi
^{n-i}.\label{Viete}%
\end{equation}

Consider the $(q,p)$-coordinates defined as follows \cite{bl04}:
\begin{equation}
q^{i}=(-1)^{i}\sigma_{i}(\lambda),\quad p_{i}=-\sum\limits_{k=1}^{n}%
(\lambda^{k})^{n-i}\mu_{k}/\Delta_{k},\quad i=1,\dots,n.\label{first}%
\end{equation}
Notice that $q^{i}$ are nothing but the coefficients of the
characteristic polynomial of $L$ (\ref{Viete}).
In the $(q,p)$-coordinates we have 
\[
(G_{0})^{rs}=\delta_{n+j}^{r+s}
+\sum\limits_{j=1}^{n-1}q^{j}\delta_{n+j}^{r+s},\quad
L_{j}^{i}=-\delta^1_j q^i+\delta^{i+1}_j,
\]
whence for $m=0,\dots, n$ we find \cite{bl05}
\[
(G_{m})^{rs}=\left\{ \ba{l}
\delta_{n-m+1}^{r+s}+\sum\limits_{j=1}^{n-m-1}q^{j}\delta_{n-m+j+1}^{r+s},\quad r,s=1,\dots,n-m,\\[5mm]
-\hspace{-5mm}\sum\limits_{j=n-m+1}^{n}q^{j}\delta_{n-m+j+1}^{r+s},\quad r,s=n-m+1,\dots,n,\\[5mm]
0\quad\mbox{otherwise.} \ea \right.
\]
Here and below $\delta_i^j$ stands for the Kronecker delta.
\looseness=-1

The geodesic Hamiltonians $E_{m,r}=\frac{1}{2}\mu^{T}K_{r}G_{m}\mu$ are polynomial
in 
the $(p,q)$-coordinates for $m=0,1,2,\dots$ \cite{bl04}. In
particular, for $m=0,\dots,n$ we have
\[
\ba{l}
E_{m,1}=\frac{1}{2}\sum\limits_{j=1}^{n-m}p_{j}p_{n-m-j+1}
+\frac{1}{2}\sum\limits_{k=1}^{n-m-1}q^{k}\sum\limits_{j=k+1}%
^{n-m}p_{j}p_{n-m+k-j+1}
-\frac{1}{2}\sum\limits_{k=1}^{m}q^{n-m+k}%
\sum\limits_{j=1}^{k}p_{n-m+j}p_{n-m+k-j+1}.\label{E1}%
\ea
\]

The basic separable potentials are given by the recursion relations
\cite{bl04}
\begin{equation}
\begin{array}{l}
\fl V_{r}^{(k+1)}=V_{r+1}^{(k)}+V_{r}^{(1)}V_{1}^{(k)},\quad
k=1,2,\dots, \quad
 V_{r}^{(1)}=-q^r, \\
\fl V_r^{(0)}=0, \\
\fl V_{r}^{(-k-1)}=V_{r-1}^{(-k)}+V_{r}^{(-1)}V_{n}^{(-k)}, \quad
k=1,2,\dots,\quad V_{r}^{(-1)}=-q^{r-1}/q^n,\label{recursion}%
\end{array}%
\end{equation}
where we tacitly assume that $q^0\equiv 1$ and that $q^i\equiv 0$
for $i>n$.

For any fixed $m$ and $k$ the Hamiltonians $H_{r}^{(m,k)}$  are in
involution with respect to the canonical Poisson bracket on
$T^{\ast}\mathcal{Q}$
\[
\{f,g\}=\sum\limits_{j=1}^{n}
{\ds
\frac{\p f}{\p \lambda^j}\frac{\p g}{\p \mu_j}
-\frac{\p g}{\p \lambda^j}\frac{\p f}{\p \mu_j}
}.
\]
These Hamiltonians are automatically separable in the
$(\lambda,\mu)$-coordinates because they satisfy the St\"{a}ckel
separation relations by construction \cite{mac2005}.
Note that the transition
from the $(\lambda,\mu)$- to the $(p,q)$-coordinates
is a canonical transformation.


\section{Superintegrability and flat coordinates}
For any natural number $n\geq 2$ let
\[
H_{m,r}\equiv E_{m,r}+\sum\limits_{k=k_0}^{k_1}c_k V_{r}^{(k)},\quad r=1,\dots,n,
\]
where $c_k$ are arbitrary constants and $m$, $k_0$ and $k_1$ are integers.
It is readily seen that these Hamiltonians are in involution for any fixed $m$, $k_0$ and $k_1$
(see Theorem~1 of \cite{aspr} for details):
\[
\{H_{m,r},H_{m,s}\}=0,\quad r,s=1,\dots,n.
\]


We have the following straightforward generalization of Theorem 1 from
\cite{bl04} (see also \cite{aspr}):
\begin{theorem}\label{parsupint}
Given a natural $n\geq 2$ and an integer $m\in\{0,\dots ,n-1\}$, let $k_0=-m$ and
$k_1=n-1-m$. Then $p_{n-m}$ is an integral of motion for $H_{m,1}$,
i.e., $\{H_{m,1},p_{n-m}\}=0$.

Moreover, $F_{m,s}=\{ H_{m,s},p_{n-m}\}$ are
additional integrals of motion for $H_{m,1}$:
$\{H_{m,1},F_{m,s}\}=\nolinebreak 0$, $s=2,\dots,n$,
and the $(2n-1)$ integrals of motion for $H_{m,1}$ ($H_{m,r}$, $r=1,\dots,n$, and
$F_{m,s}$, $s=2,\dots,n$) are functionally independent, so $H_{m,1}$
is maximally superintegrable.
\end{theorem}

Under the assumptions of Theorem~\ref{parsupint} the Hamiltonian $H_{m,1}$
involves $n-1$ parameters $c_i$ (note that $V_1^{(0)}\equiv 0$).

\begin{prop}[\cite{bl05}]
Given a natural $n\geq 2$ and an integer $m\in\{0,\dots,n\}$,
the metric $G_{m}$ in the coordinates $r^i$ defined by
the formulas
\begin{equation}\label{flat}
\begin{array}{ll}
q^{i}  &
=r^{i}+{\frac{1}{4}}\sum\limits_{j=1}^{i-1}r^{j}r^{i-j},\quad
i=1,\dots,n-m,\nonumber\\
q^{i}  &  =-{\frac{1}{4}}\sum\limits_{j=i}^{n}r^{j}r^{n-j+i},\quad
i=n-m+1,\dots,n,\label{second}
\end{array}
\end{equation}
takes the form
\begin{equation}\label{gmtr}
(G_{m})^{kl}=\left(
\delta_{n-m+1}^{k+l}+\delta_{2n-m+1}^{k+l}\right).
\end{equation}
The transition from $(q,p)$- to $(r,\pi)$-coordinates, where
$\pi_{k}=\sum\limits_{i=1}^{n}\displaystyle\frac{\partial q^{i}}{\partial r^{k}}p_{i}$, $k=1,\dots,n$,
is a canonical transformation,
and we have
\begin{equation}
E_{m,1}\equiv H^{(0)}_{m,1}=
\frac{1}{2}\left(\sum\limits_{j=1}^{n-m}\pi_{j}\pi_{n-m+1-j}%
+\sum\limits_{j=n-m+1}^{n}\pi_{j}\pi_{2n-m+1-j}\right). \label{E2}%
\end{equation}
The tensor $L$ in the coordinates $r^i$ takes the form:
\[
\ba{l}
\fl\mbox{for $m<n$:}\quad L^i_j=\delta^{i+1}_j(1-\delta^i_{n-m}) -\frac12 r^i \delta^1_j
-\frac12 r^{n-j-m+1+n[(j+m-1)/n]} \delta^i_{n-m}\\[5mm]
\fl\mbox{for $m=n$:}\quad L^i_j=\delta^{i+1}_j+\frac14 r^i
r^{n-j+1}. \ea
\]
\end{prop}
Here and below $[k]$ denotes the largest integer less than or equal to $k$.

The canonical coordinates $(r,\pi)$ are
not orthogonal, but the metric tensor $G_{m}$ is constant in these
coordinates. Bringing $G_m$ into the canonical form, with $+1$
and $-1$ at the diagonal and zeros off the diagonal,
is possible \cite{bl05} but we shall not need this here.


Recall \cite{bl04} that for $k=1,\dots, n-1$ the potentials
$V_{1}^{(k)}$ are independent of $q^{j}$ with $j=k+1,\dots,n$.
Likewise, for $k=1,\dots, n-1$ the potentials
$V_{1}^{(-k)}$ are independent of $q^{j}$ with $j=1,\dots,n-k$.
On the other hand, the change of variables (\ref{flat}) is partially
triangular: $q^i$ with $i=1,\dots,n-m$ depend only on $r^1,\dots,
r^i$ while
$q^i$ with $i=n-m+1,\dots,n$ depend only on 
$r^i,\dots, r^n$.
\looseness=-1

Hence the coordinates $r^i$ enjoy the following
remarkable property:
\begin{theorem}\label{cyclcor}Given a natural\ $n\geq 2$,
and two non-negative integers, $m\in\{0,\dots,n-2\}$ and $k$,
consider the Hamiltonians $H_{m,1}^{(k,+)}=E_{m,1}+\sum\limits_{j=1}^k c_j
V_{1}^{(j)}$ and $H_{m,1}^{(-k,-)}=E_{m,1}
+\sum\limits_{j=1}^k c_j V_{1}^{(-j)}$,
where $c_j$ are arbitrary constants.
\looseness=-1

If $k\in\{1,\dots,n-m\}$
then the Hamiltonian $H_{m,1}^{(k,+)}$ commutes not only
with the `standard' integrals
$H_{m,r}^{(k,+)}=E_{m,r}+\sum\limits_{j=1}^k c_j V_{r}^{(j)}$,
$r=2,\dots,n$, but also with $\pi_j$, $j=k+1,\dots,n$, i.e., $r^j$
are cyclic variables for $H_{m,1}^{(k,+)}$ for $j=k+1,\dots,n$.

Likewise, if $k\in\{0,\dots,m\}$ then the Hamiltonian $H_{m,1}^{(-k,-)}$
commutes, in addition to
$H_{m,r}^{(-k,-)}= E_{m,r}+\sum\limits_{j=1}^k c_j V_{r}^{(-j)}$,
$r=2,\dots,n$,
with
$\pi_j$, $j=1,\dots,n-k$, i.e., $r^j$ are
cyclic variables for $H_{m,1}^{(-k,-)}$ for $j=1,\dots,n-k$.
\end{theorem}

Note that, in contrast with the above,
the additional integrals of motion for $H_{i,1}^{(k)}$
found earlier in \cite{bl04} were quadratic rather than linear in
momenta.


\section{
Exact solvability in $n$ dimensions}
Because of the special form (\ref{gmtr}) of the metric $G_m$ in the
variables $r^1,\dots,r^n$ the existence of cyclic variables
simplifies solving
the Hamilton--Jacobi and the Schr\"odinger equations for the
Hamiltonians $H_{m,1}^{(k,+)}$ with $m=0,\dots,n-2$ and
$k=1,\dots,n-m-1$ and $H_{m,1}^{(-k,-)}$ for $m=0,\dots,n-2$
and $k=0,\dots,m$
even more than one could expect, {\em especially} provided $n$ is
sufficiently large.

Let
$\mathcal{P}_s\equiv-\rmi \partial/\partial r^s$, $s=1,\dots,n$,
where $\rmi=\sqrt{-1}$.
The following results are readily verified by straightforward computation:

\begin{theorem}\label{qposth}
Fix a non-negative integer $m$, a natural $k$,
and $k$ constants $c_j$, $j=1,\dots,k$.
Then for any natural $n$ such that $n\geq 2k+m$
the most general common eigenfunction $\psi$ of
\looseness=-1
\[
\fl\mathcal{H}^{(k,+)}_{m,1}\equiv
\frac{1}{2}\left(\sum\limits_{a=1}^{n-m} \mathcal{P}_a
\mathcal{P}_{n-m+1-a}+\sum\limits_{b=n-m+1}^{n}
\mathcal{P}_b \mathcal{P}_{2n-m+1-b}\right)
+\sum\limits_{j=1}^{k} c_j V_{1}^{(j)}
\]
and of $\mathcal{P}_j$, $j=k+1,\dots,n$,
with the
eigenvalues $E$ and $\pi_j$, $j=k+1,\dots,n$, respectively, is
quasiclassical and reads
$\psi=\exp(\rmi S)$,
where
$S(r^1,\dots,r^n)=S_0(r^1,\dots,r^k)+\sum\limits_{j=k+1}^n r^j \pi_j$
satisfies the stationary Hamilton--Jacobi equation for ${H}^{(k,+)}_{m,1}$,
and $S_0$ is a general solution of the reduced Hamilton--Jacobi equation,
a {\em first
order linear} PDE in $k$ independent variables: 
\be\label{hj} \ba{l}
\fl\sum\limits_{j=1}^{k} \pi_{n-m+1-j}
{\displaystyle\frac{\p S_0}{\p r^j}}+\sum\limits_{j=1}^{k} c_j
V_{1}^{(j)} =\varepsilon,
\ea \ee
where
\[
\fl\varepsilon=
E- \hspace{-3mm}
\sum\limits_{j=k+1}^{[(n-m)/2]} \hspace{-3mm}\pi_{n-m+1-j} \pi_j
-\frac{(n-m-2[(n-m)/2])}{2}\pi_{n-m-[(n-m)/2]}^2
-\frac12 \sum\limits_{j=n-m+1}^{n} \hspace{-3mm}\pi_j \pi_{2n-m+1-j}.
\]
\end{theorem}

\begin{theorem}\label{negkth}
Fix a natural $k$, a non-negative integer $m\geq 2k$,
and $k$ constants $c_j$, $j=1,\dots,k$.
Then for any natural $n\geq m$
the most general common eigenfunction $\psi$ of
\[
\fl\mathcal{H}^{(-k,-)}_{m,1}\equiv
\frac{1}{2}\left(\sum\limits_{a=1}^{n-m} \mathcal{P}_a
\mathcal{P}_{n-m+1-a}+\sum\limits_{b=n-m+1}^{n}
\mathcal{P}_b \mathcal{P}_{2n-m+1-b}\right)
+\sum\limits_{j=1}^{k} c_j V_{1}^{(-j)}
\]
and of $\mathcal{P}_j$, $j=1,\dots,n-k$,
with the
eigenvalues $E$ and $\pi_j$, $j=1,\dots,n-k$, respectively, is
quasiclassical and reads
$\psi=\exp(\rmi S)$,
where
$S(r^1,\dots,r^n)=S_0(r^{n-k+1},\dots,r^n)
+\sum\limits_{j=1}^{n-k} r^j \pi_j$
satisfies the stationary Hamilton--Jacobi equation for $H^{(-k,-)}_{m,1}$,
and $S_0$ is a general solution of the reduced Hamilton--Jacobi equation,
a {\em first
order linear} PDE in $k$ independent variables: 
\be\label{hj2} \ba{l} \fl \sum\limits_{j=n-k+1}^{n} \pi_{2n-m+1-j}
{\displaystyle\frac{\p S_0}{\p r^j}}+\sum\limits_{j=1}^{k} c_j
V_{1}^{(-j)} =\tilde\varepsilon, \ea \ee where
\[
\fl\tilde\varepsilon=E- {\displaystyle\frac{1}{2}}
\sum\limits_{j=1}^{n-m} \pi_{n-m+1-j} \pi_j
-\sum\limits_{j=n-[m/2]+1}^{n-k} \pi_j \pi_{2n-m+1-j}-
\frac{(m-2[m/2])}{2}\pi_{n-[m/2]}^2.
\]
\end{theorem}

Let us briefly outline the integration strategy for (\ref{hj})
and (\ref{hj2}). Consider (\ref{hj}) first.
Assume that we have found new coordinates
$z^1(r^1,\dots,r^k),\dots,z^k(r^1,\dots,r^k)$ 
such that
\[
\sum\limits_{j=1}^{k} \pi_{n-m+1-j}
{\displaystyle\frac{\p}{\p r^j}} ={\displaystyle\frac{\p}{\p z^k}}.
\]
This is always possible, but the choice of $z$'s is not unique and
depends on the particular values of $\pi_j$. Now (\ref{hj})
becomes an {\em ODE} in $z^k$ involving 
$z^1,\dots,z^{k-1}$ as parameters:
\be\label{hja} \ba{l}
\fl{\displaystyle\frac{\p S_0}{\p z^k}}+\sum\limits_{j=1}^{k} c_j
V_{1}^{(j)} =\varepsilon, \ea \ee
where $V_{1}^{(j)}$ are now
considered as functions of $z$'s.

The general solution of (\ref{hja}) reads
\be\label{gshj} \ba{l} \fl
S_0=K(z^1,\dots,z^{k-1})-\sum\limits_{j=1}^{k} \int c_j V_{1}^{(j)}dz^k
+\varepsilon z^k,
\ea \ee
where $K$ is an arbitrary
smooth function of its arguments.

Likewise, assume that we have found new coordinates $\tilde
z^{n-k+1}(r^{n-k+1},\dots,r^n),\dots,\tilde
z^{n}(r^{n-k+1},\dots,r^n)$ such that
\[
\sum\limits_{j=n-k+1}^{n} \pi_{2n-m+1-j} {\displaystyle\frac{\p}{\p
r^j}}={\displaystyle\frac{\p}{\p\tilde  z^n}}.
\]
Then the general solution of (\ref{hj2}) reads \be\label{gshj2}
\ba{l} \fl S_0=\tilde K(\tilde z^{n-k+1},\dots,\tilde z^{n-1})
-\sum\limits_{j=1}^{k} \int c_j V_{1}^{(-j)}d \tilde z^n
+\tilde\varepsilon \tilde z^n, \ea \ee where $\tilde K$ is an
arbitrary smooth function of its arguments.

Thus, the stationary Schr\"odinger equations for the Hamiltonians
$H^{(k,+)}_{m,1}$ and $H^{(-k,-)}_{m,1}$ with $m=0,\dots,n-2$
(for $H^{(-k,-)}_{m,1}$ we have an extra condition $m\geq 2k$)
and arbitrary constants
$c_i$, $i=1,\dots,k$, in the
space of $n$ dimensions for {\em any} $n\geq 2k+m$ essentially
reduce to a
linear {\em first-order} PDEs in $k$ independent
variables, and these PDE can be explicitly solved. This is a rather
surprising result, as only a few potentials for which the Schr\"odinger equation
is exactly solvable in the space of arbitrarily high dimension were known so
far, cf.\ the discussion in Introduction.
\looseness=-1


It is readily seen that there exists no real value
of $E$ for which the eigenfunction
$$
\psi=\exp\left(\rmi \left(S_0
+\sum\limits_{j=k+1}^{n} \pi_j r^j\right)\right)$$ of
$\mathcal{H}_{m,1}^{(k,+)}$
constructed in Theorem~\ref{qposth}
with $S_0$ given by (\ref{gshj}) can have a finite norm
\[
\int_\mathcal{Q}|\psi|^2 dr^1\dots dr^n\sim
\int_\mathcal{Q}|\exp(K(z^1,\dots,z^{k-1}))|^2 dz^1\dots dz^n.
\]
The latter integral obviously diverges for any choice
of $K$ because of integration over $z^{k+1}$,\dots, $z^n$.
Therefore no common eigenfunction $\psi$ of $\mathcal{H}^{(k,+)}_{m,1}$ and $\mathcal{P}_j$,
$j=k+1,\dots,n$, can belong to the discrete
spectrum of $\mathcal{H}^{(+)}_{m,1}$.
%
In a similar fashion we can show that no common eigenfunction
$\psi$ of $\mathcal{H}_{m,1}^{(-k,-)}$ and $\mathcal{P}_j$,
$j=1,\dots,n-k$, can belong to the discrete
spectrum of $\mathcal{H}_{m,1}^{(-k,-)}$.

\section{Example: an anharmonic oscillator}

Consider the Hamiltonian
\[
H\equiv H_{0,1}^{(3)}=\frac12\sum\limits_{k=0}^{n-1} q^{k} \sum
\limits_{j=k+1}^{n}p_{j}p_{n+k-j+1}-q^3+2 q^1 q^2 - (q^1)^3,
\]
in the $(q,p)$-cootdinates.
It commutes with
\[
H_{i}\equiv H_{0,i}^{(3)}=\frac12\sum\limits_{i,j=1}^{n} (K_r G)^{ij}p_i p_j-q^{i+2}
+q^{i+1}q^1+q^{i}q^2-q^i (q^1)^2,\quad i=2,\dots,n,
\]
by construction.

\subsection{Superintegrability}
The Hamiltonian $H$ is superintegrable  \cite{bl04}
because it also
commutes with the function
$$
I=\left\{
\ba{l}\frac12 p_2^2+(q^1)^2,\quad n=2,\\
\frac12 p_3^2-q^1,\quad n=3,\\
p_n,\quad n\geq 4,
\ea
\right.
$$
and hence $F_{r}=\{I, H_r\}$, $r=2,\dots,n$
also Poisson commute with $H$, and all functions in the set
$\{ H_r, r=1,\dots,n, F_s, s=2,\dots,n\}$
are functionally independent.

For $n\geq 4$ the additional integrals $F_r$ are simply
$F_{r}=\{p_n, H_r\}=-\p H_r/\p q^n$,
and we readily find that 
$$
F_r=\frac12 \sum\limits_{i=n-r+2}^{n}
\sum\limits_{j=2n+2-i-r}^{n}q^{i+j+r-2n-2}p_i p_j
+\delta^{r+2}_n-\delta^{r+1}_n q^1-\delta^{r}_n (q^2 -(q^1)^2).
$$
In particular, we obtain
\be\label{f2} F_2=p_n^2/2,
\ee
so $F_2$ is simply a
half of square of $I$, and
it is straightforward to verify 
(see Theorem~5 of \cite{aspr} for
a more general result of this kind)
that we have
$$
\{ p_n, F_r\}=0,\quad r=2,\dots,n,
\quad
\{F_r, F_s\}=0,\quad r,s=2,\dots,n.
$$


It can be readily inferred from the above and from Theorem \ref{parsupint}
that the quantities $\{H, I, F_3, \dots, F_n\}$ are functionally independent
and Poisson commute for all $n\geq 2$.

Thus, the Hamiltonian $H$ is maximally
superintegrable for all $n=2,3,\dots$. For $n=3$ we have, in
addition to $H$, $H_2$, $H_3$ and $I$, the following integral
$K=p_3 p_2 + q^1 p_3^2/2 -q^2+(q^1)^2=\pi_2 \pi_3 -r^2+3(r^1)^2/4$
which is quadratic in momenta.
Note that $I$ and $K$ commute.

\subsection{Quantization and solution of equations of motion\\ in the
$(r,\pi)$-coordinates}

In the
$(r,\pi)$-coordinates we have
\[
H=\frac12 \sum\limits_{j=1}^n
\pi_j\pi_{n+1-j}-r^3+\frac32 r^1 r^2 -\frac12 (r^1)^3.
\]
The quantization is obvious: $H$ goes into the operator
\[
\mathcal{H}=\frac12 \sum\limits_{j=1}^n
\mathcal{P}_j\mathcal{P}_{n+1-j}-r^3+\frac32 r^1 r^2 -\frac12 (r^1)^3.
\]

For $n\geq 4$ we can look for
the common eigenfunctions $\psi$ of $\mathcal{H}$ and $\mathcal{P}_i$, $i=4,\dots,n$:
\be\label{eig}
\mathcal{H}\psi =E\psi,\quad \mathcal{P}_i\psi=\pi_i\psi,\quad i=4,\dots,n.
\ee
By Proposition 2, for $n\geq 6$ finding $\psi$
requires solving the equation (\ref{hj}) that becomes
\be\label{hj3}
\ba{l}
\sum\limits_{k=1}^{3} \pi_{n+1-k}\p S_0/\p r^{k}
-r^3+\frac32 r^1 r^2 -\frac12 (r^1)^3
=\varepsilon,
\ea
\ee
where
\[
\varepsilon=E-\sum\limits_{k=4}^{[n/2]}
\pi_{n+1-k}\pi_{k}-\frac{(n-2[n/2])}{2}\pi_{[n/2]+1}^2.
\]
Eq.(\ref{hj3}) is a {\em first} order PDE
that can be readily solved in full generality.

Let 
$K(\omega_1,\omega_2)$ stand below for an arbitrary (smooth)
function of its arguments.

If $\pi_n\neq 0$ then the general solution of (\ref{hj3}) reads
\[
\ba{l}
\fl
S_0=\ds\frac{(r^1)^4}{8 \pi_n}+\frac{\pi_{n-1}(r^1)^3}{4 \pi_n^2}
+\left(-\frac{3 r^2}{4 \pi_n}
-\frac{\pi_{n-2}}{2 \pi_n^2}\right) (r^1)^2
+\frac{(r^3+\varepsilon)r^1}{\pi_n}\\[5mm]
+\ds K\left(\frac{(r^2 \pi_n-\pi_{n-1}r^1)}{\pi_n},
\frac{(r^3\pi_n-\pi_{n-2}r^1)}{\pi_n}\right). \ea
\]

If $\pi_n=0$ but $\pi_{n-1}\neq 0$ then the general solution of (\ref{hj3}) is
\[
\ba{l} \fl S_0=\ds \frac{(r^1)^3 r^2}{2 \pi_{n-1}} -\frac{3
r^1(r^2)^2}{4 \pi_{n-1}} -\frac{\pi_{n-2} (r^2)^2}{2 \pi_{n-1}^2}+
\frac{(r^3+\varepsilon)r^2}{\pi_{n-1}}
+\ds K\left(r^1, \frac{(r^3\pi_{n-1}-\pi_{n-2}
r^2)}{\pi_{n-1}}\right). \ea
\]

Finally, if $\pi_n=\pi_{n-1}=0$ but $\pi_{n-2}\neq 0$ then
the general solution of (\ref{hj3}) has the form
\[
S_0=\ds\frac{(r^3)^2-3 r^1 r^2 r^3+(r^1)^3 r^3
+2\varepsilon r^3}{2 \pi_{n-2}}+K(r^1, r^2).
\]

By Proposition 2 for $n\geq 6$ the most general
common eigenfunction
$\psi$ of $\mathcal{H}$ and of $\mathcal{P}_i$, $i=4,\dots,n$
is
\[
\psi=\exp\left(\rmi \left(S_0
+\sum\limits_{j=4}^n \pi_j r^j\right)\right)
\]
where $S_0$ is given above. This eigenfunction
is {\em not} square integrable for any choice of $K$
and for any values of $\pi_i$.

An integral of the stationary Hamilton--Jacobi equation for $H$ is
given by
\[
S=S_0+\sum\limits_{j=4}^n \pi_j r^j.
\]

\subsection{Special cases: $n\leq 5$}
We list below the exact solutions for the corresponding Schr\"odinger and Hamilton--Jacobi equations.
None of the eigenfunctions listed below is square integrable.
\subsubsection*{Case 1: $n=2$}

Here $q^3=0$ and there is a pair
of commuting operators $\mathcal{H}$ and $\mathcal{I}$.
In the $(r,\pi)$-coordinates\\ we have
\looseness=-1
\[
\mathcal{H}=\mathcal{P}_1\mathcal{P}_2
+  2 r^1 r^2-(r^1)^3/2,\quad
\mathcal{I}=\mathcal{P}_2^2/2+(r^1)^2.
\]
and their common eigenfunction $\psi$ satisfying
$\mathcal{H}\psi=E\psi$, $\mathcal{I}\psi=\lambda_1\psi$
has, up to multiplication by an arbitrary constant, the form
$$
\psi={\ds\frac{1}{\sqrt{(r^1)^2-\lambda_1}}}
\left(r^1+\sqrt{(r^1)^2-\lambda_1}\right)^{E/\sqrt{2}}
\exp\left(\sqrt{2((r^1)^2-\lambda_1)}(r^2-(r^1)^2/12-\lambda_1/6)\right)
$$
We actually have two such eigenfunctions, as
$\sqrt{(r^1)^2-\lambda_1}$ is a two-valued function.

The eigenfunction $\psi$ is not quasiclassical: we have a complete
integral of the stationary Hamilton--Jacobi equation for $H$ of the
form
\[
\ba{l}
\fl S=-\rmi\ln\psi+\ds\frac{\rmi}{3\sqrt{2}}((r^1)^2-\lambda_1)^{1/2}(-2\lambda_1-(r^1)^2+12 r^2)
-\ds\frac{\rmi}{2}\ln((r^1)^2-\lambda_1),
\ea
\]
or equivalently,
\[
\fl S=\ds\frac{E}{\sqrt{2}}\arctan\left(\ds\frac{r^1}{((r^1)^2-\lambda_1)^{1/2}}\right)
-\ds\frac{1}{12}(12 r^2-(r^1)^2-2\lambda_1)(2(\lambda_1-(r^1)^2))^{1/2}.
\]

\subsubsection*{Case 2: $n=3$}
Now we have a triplet of commuting operators $\mathcal{H}$,
$\mathcal{I}=\pi_3^2/2-r^1$, and
$\mathcal{F}_2=\mathcal{P}_3^3/2-3 r^1\mathcal{P}_3/2-\mathcal{P}_2$
and their common eigenfunction $\psi$ such that $\mathcal{H}\psi=E\psi$,
$\mathcal{I}\psi=\lambda_1\psi$ and $\mathcal{F}_2\psi=\lambda_2 \psi$
reads
$$
\ba{l} \psi={\ds\frac{1}{(r^1+\lambda_1)^{1/2}}}
\exp\biggl(\rmi\sqrt{2}(\lambda_1+r^1)^{1/2} r^3
+\rmi (-\lambda_2 + \sqrt{2}(r^1 +\lambda_1)^{1/2} \lambda_1) r^2 \\[5mm]
-\ds\frac{\rmi}{\sqrt{2}}(r^1 +\lambda_1 )^{1/2} r^1 r^2
+\frac{\rmi}{28}\sqrt{2}(r^1 +\lambda_1 )^{1/2} (r^1)^3
+\frac{\rmi}{4} (-\lambda_2+\frac{3}{7}
\sqrt{2}(r^1 +\lambda_1 )^{1/2} \lambda_1) (r^1)^2\\[5mm]
-\ds\rmi (-\lambda_2 + \frac{\sqrt{2}}{7}(r^1 +\lambda_1
)^{1/2}\lambda_1) \lambda_1  r^1 +\frac{\rmi\sqrt{2}}{14}(r^1
+\lambda_1 )^{1/2} (-10 \lambda_1 ^3+ 14 E-7 \lambda_2 ^2)\biggr)
\ea
$$

If we take another triple of commuting operators $\mathcal{H}$,
$\mathcal{I}=\pi_3^2/2-r^1$, and $\mathcal{K}=\mathcal{P}_2
\mathcal{P}_3 -r^2+3(r^1)^2/4$ then their common eigenfunction $\psi$
such that $\mathcal{H}\psi=E\psi$, $\mathcal{I}\psi=\lambda_1\psi$
and $\mathcal{K}\psi=\nu\psi$ reads
\looseness=-1
\[
\ba{l}
\psi={\ds\frac{1}{(r^1+\lambda_1)^{3/4}}}
\exp\biggl(\rmi\sqrt{2}(\lambda_1+r^1)^{1/2} r^3
-\ds\frac{\rmi}{\sqrt{2}(\lambda_1+r^1)^{1/2}}\biggl(
-\frac{23}{224}(r^1)^4 \\[5mm]
-\ds\frac{\lambda_1}{28}(r^1)^3
+\frac34 r^2 (r^1)^2
\ds
+\frac{1}{28}(-7\nu+2\lambda_1^2) (r^1)^2
-\frac17(14 E+2\lambda_1^3-7\lambda_1\nu)r^1
-(r^2)^2/2 -\nu r^2 \\[5mm]
-(-2\lambda_1^2\nu+\frac47 \lambda_1^4
+2\lambda_1 E+\frac12\nu^2)\biggr)\biggr). \ea
\]

In the latter case the eigenfunction $\psi$ is almost quasiclassical:
we have a complete integral of the Hamilton--Jacobi equation for $H$
of the form
\[
S=-\rmi\ln\psi+\ds\frac{3\rmi}{4}\ln(r^1+\lambda^1).
\]

In both cases we again actually have two eigenfunctions as
$(r^1+\lambda_1)^{1/2}$ is two-valued.

\subsubsection*{Case 3: $n=4$}
The common eigenfunction of $\mathcal{P}_4$, $\mathcal{H}$ and
$\mathcal{F}_i$, $i=2,3$, with the respective eigenvalues $\pi_4$,
$E$, $\lambda_2$,$\lambda_3$ reads
\[
\ba{l} \fl \ds\psi=\exp\biggl(\rmi \pi_4 r^4+ \frac{\rmi}{16\pi_4^5}
(2 \pi_4^4+2+3 \pi_4^2) (r^1)^4 +\frac{\rmi}{4 \pi_4^5} \lambda_2
(\pi_4^2+2) (r^1)^3-\frac{\rmi}{4 \pi_4^3} (3 \pi_4^2+2) (r^1)^2
r^2\\[5mm]
\fl\ds -\frac{\rmi}{4 \pi_4^5} (-3 \lambda_2 ^2+2 \pi_4^2 \lambda_3
)(r^1) ^2 -\frac{\rmi \lambda_2}{\pi_4^3}r^2 r^1 +\frac{\rmi}{\pi_4}
r^3 r^1 +\frac{\rmi}{2 \pi_4^5} (-2 \lambda_2 \pi_4^2 \lambda_3
+\lambda_2 ^3+2 \pi_4^4 E)r^1\\[5mm]
\fl\ds+\frac{\rmi}{2\pi_4} (r^2)^2 +\frac{\rmi}{2\pi_4^3}
(-\lambda_2^2+2 \pi_4^2 \lambda_3 ) r^2 +\frac{\rmi
r^3}{\pi_4}\lambda_2\biggr)
\ea
\]

This eigenfunction is quasiclassical: $S=-\rmi\ln \psi$ is a
complete integral for the stationary Hamilton--Jacobi equation for
$H$.

\subsubsection*{Case 4: $n=5$}The common eigenfunction of $\mathcal{P}_i$,
$i=4,5$, $\mathcal{H}$ and $\mathcal{F}_i$, $i=3,4$, with the
respective eigenvalues $\pi_4,\pi_5$, $E$, $\lambda_3$,$\lambda_4$
reads, up to an overall constant factor,
\[
\ba{l} \fl\psi=\ds\exp\biggl(\rmi \pi_5 r^5+\rmi \pi_4
r^4+\frac{\rmi}{8 \pi_5} (r^1)^4+\frac{\rmi}{12\pi_5^3} (3 \pi_5
\pi_4+2)(r^1)^3
-\frac{3\rmi}{4\pi_5} (r^1)^2 r^2\\[5mm]
+\ds\frac{\rmi}{4\pi_5^3} (2 \lambda_3 +\pi_4^2)(r^1)^2
-\frac{\rmi}{\pi_5^2} \pi_4 r^2 r^1
+\frac{\rmi}{\pi_5} r^3 r^1\\[5mm]
-\ds\frac{\rmi}{8 \pi_5^3} (-8 E \pi_5^2-4 \lambda_3^2+3 \pi_4^4+8
\pi_4 \lambda_4 \pi_5-4 \pi_4^2 \lambda_3 )r^1 +\frac{\rmi}{2\pi_5}
(r^2)^2\\[5mm]
\ds +\frac{\rmi}{2\pi_5^2} (2 \lambda_4
 \pi_5-2 \pi_4 \lambda_3 +\pi_4^3) r^2
-\frac{\rmi}{2\pi_5} (\pi_4^2-2 \lambda_3 ) r^3\biggr) \ea
\]

Unlike the previous case, this wave function is not quasiclassical:
there is a complete integral of the stationary Hamilton--Jacobi
equation for $H$ of the form
\[
S=-\rmi\ln\psi +\frac{1}{24\pi_5^3}(\pi_4^2-2r^1-2\lambda_3)^3
\]


\subsection*{Acknowledgements}

It is my great pleasure to thank Prof.\ Pavel Winternitz
for highly stimulating discussions on the subject
of the present work and for the encouragement. I am also very
pleased to thank Prof.\ Maciej B\l aszak for useful suggestions.

This research was supported in part by the Czech Grant Agency (GA%
\v{C}R) under grant No.\ 201/04/0538 and by the Ministry of
Education, Youth and Sports of the Czech Republic under grant MSM
4781305904
and the development project No. 254/b for the year 2004.
The author is pleased to acknowledge the warm hospitality of Centre
de Recherches Math\'ematiques, Universit\'e de Montr\'e{}al, where
the present work was initiated, and especially of Prof.\ Pavel
Winternitz.
\looseness=-1


\end{document}